\documentclass[pra,
twocolumn,
floatfix]{revtex4-1}

\usepackage[utf8]{inputenc}
\usepackage{color}
\usepackage{graphicx}
\usepackage{lmodern}
\usepackage{amsmath}
\usepackage{amssymb}
\usepackage{mathtools}
\usepackage{capt-of}
\usepackage{hyperref}

\def\beq{\begin{equation}}
\def\eeq{\end{equation}}

\begin{document}
\title{Two-color phase-of-the-phase spectroscopy with circularly polarized laser pulses}
\author{V. Tulsky, M. A. Almajid, and D. Bauer}
\affiliation{Institute of Physics, University of Rostock, 18051 Rostock, Germany}
\date{\today}
\begin{abstract}
Phase-of-the-phase spectroscopy using two-color colinearly polarized laser pulses  has been introduced and experimentally applied to strong-field tunneling ionization in S. Skruszewicz  {\em et al}.,  Phys. Rev. Lett. {\bf115}, 043001 (2015) and recently to multiphoton ionization in M. A. Almajid  {\em et al}., J. Phys. B: At. Mol. Opt. Phys. {\bf 50}, 194001 (2017). The idea behind  phase-of-the-phase spectroscopy is to study in a systematic way the change in the photoelectron yield as a function of the relative phase between the strong fundamental field component (of carrier frequency $\omega$) and a weak, second color component (e.g., $2\omega$). The observable of interest is the photoelectron-momentum-dependent phase of the change in the electron yield with respect to the relative phase, hence the name ``phase of the phase.''   In the present paper,  phase-of-the-phase spectroscopy is extended to circularly polarized light. With a small, counter-rotating $2\omega$-component,  photoelectron spectra  have a three-fold symmetry in the polarization plane. The same is true for the corresponding phase-of-the-phase spectra. However, a peculiar, very sharp phase-flip by $\pi$ occurs at a certain radial momentum  of the photoelectron that is sensitive to both laser parameters and the ionization potential.  Results from the numerical solution of the time-dependent Schr\"odinger equation are compared to those from the strong-field approximation. An analytical expression for the momentum at which the phase-of-the-phase flipping occurs is presented. 
\end{abstract}

\maketitle

\section{INTRODUCTION}
 Bichromatic laser fields are of interest because they offer additional control over the generation of high-order harmonics (HHG), terahertz (THz) radiation, or shaping of photoelectron wave-packets (see, e.g., \cite{Miloshevic_PhysRevA_2000, Cook_OptLett_2000, Kim_ApplPhysB_2004, Kramo_LasPhysLett_2007, Murray_JPhysB_2010, Vvedenskii_PhysRevLett_2014, Zhang_PhysRevA_2014, Mancuso_PhysRevA_2015, Richter_PhysRevLett_2015, Meng_ApplPhysLett_2016, Habibovic_OptQuantEl_2018, Yu_ChemPhysLett_2018} and references therein). Recently, the momentum-resolved ionization probability of atoms in a strong linearly polarized laser field of frequency $\omega$ plus an additional component of doubled frequency $2\omega$ was investigated with respect to its dependence on the relative phase between the two field components \cite{Skruszewicz_PhysRevLett_2015, Eicke_JPhysB_2017, Almajid_JPhysB_2017, Wurzler_JPhysB_2018,  Porat_NatComm_2018, Tan_OptQuantEl_2018}. The method has been named ``phase-of-the-phase'' (PP) spectroscopy in \cite{Skruszewicz_PhysRevLett_2015, Almajid_JPhysB_2017, Wurzler_JPhysB_2018, Tan_OptQuantEl_2018} because the general idea is as follows. Instead of considering many photoelectron spectra (PES) for the various possible relative phase shifts between $\omega$ and $2\omega$ component of the laser field,  the {\em change} of the PES as a function of $\varphi$ is the basic observable in PP spectroscopy. Since the laser field and thus the PES is periodic in $\varphi$, a Fourier transform of a sequence of PES with respect to $\varphi$ allows to condense valuable information in a few functions. Assuming the $2\omega$ field to be a small addition to the main component, one may truncate the Fourier series of the PES $Y({\bf p}, \varphi)$ and focus on the first two terms only,
\begin{equation}
Y({\bf p}, \varphi)\simeq Y_0({\bf p}) +  \triangle Y_1({\bf p}) \cos[\varphi + \Phi_1({\bf p})].
\end{equation}
Hence, in first order the change of the PES as a function of $\varphi$ is captured just in two functions: the absolute value of the change $\Delta Y_1({\bf p})$ (relative-phase contrast) and the eponymous phase of the phase $\Phi_1({\bf p})$. Prominent features appearing at certain momenta in $\Phi_1({\bf p})$ \cite{Skruszewicz_PhysRevLett_2015, Almajid_JPhysB_2017} are laser and target-sensitive and can provide more information than an ordinary PES for just the $\omega$ field. 
Since any dependence on the relative phase $\varphi$ will be spoiled by incoherent emission processes (such as thermal emission or multiple scattering in complex many-body targets), PP spectroscopy also provides the possibility to reveal coherent features when ordinary PES are cluttered by insipidly incoherent, Maxwellian contributions.
  
 In the present paper, we introduce PP spectroscopy for the generic case of a circularly polarized laser field of frequency $\omega$ and a small, counter-rotating component of frequency  $2\omega$. We observe a sharp $\pi$ phase flip in the PP spectra at a certain radial momentum and derive it analytically.  The paper is organized as follows: in Section~\ref{theory}, PP spectra for two-color counter rotating pulses are analytically and numerically studied via the strong-field approximation and the $\textit{ab initio}$ solution to the time-dependent Schr\"odinger equation. Section~\ref{concl} contains Conclusion and Outlook. The dipole approximation is well valid for the laser parameters of interest and applied throughout the paper. Atomic units are used unless stated otherwise.

\section{THEORY AND RESULTS} \label{theory}
The two-color laser field with counter rotating $\omega$ and $2\omega$ components can be represented as ${\bf E}(t)=-\partial_t{\bf A}(t)$ with a vector potential 
\begin{align}\label{A}
{\bf A}(t) &= {\bf A}_\omega(t)  + {\bf A}_{2\omega}(t) \nonumber\\ 
& = A_0 f(t)\Bigg[\bigg(\cos\omega t +\xi \cos(2\omega t + \varphi)\bigg){\bf e}_x  \nonumber \\
& \quad +  \bigg(\sin\omega t- \xi\sin(2\omega t + \varphi)\bigg){\bf e}_y\Bigg] .
\end{align}
Here, $f(t)$ is the envelope of the laser pulse, $\xi= \max|{\bf A}_{2\omega}|/\max|{\bf A}_\omega|$ is the ratio of the vector potential amplitudes, and $\varphi$ is the relative phase shift between $\omega$ and $2\omega$ component. Note that the ratio of the electric field amplitudes is $2\xi$ and the ratio of the intensities is $4\xi^2$.

For our analytical investigations we apply the strong-field approximation (SFA) \cite{Keldysh_1965, Faisal_1973, Reiss_1980}. The essential idea underlying the SFA is the assumption that, after a photoelectron has been emitted into the continuum, only the interaction between electron and the external intense laser field matters, while the influence of the binding potential on the emitted electron can be neglected. Rescattering of the photoelectron is not important for the considered circularly polarized, bichromatic laser fields with small $2\omega$ component. Hence we can restrict ourselves to the so-called direct SFA matrix element $M({\bf p})$ \cite{Miloshevich_JPhysB_2006} throughout our analytical calculations.  The photoelectron momentum distribution then reads
\begin {equation}\label{PES}
Y({\bf p},\varphi) = \vert M({\bf p},\varphi)\vert^2
\end {equation}
with \cite{Miloshevich_JPhysB_2006, Miloshevich_JModOpt_2006, Popruzhenko_JPhysB_2014}
\begin {equation}\label{Mpphi}
M({\bf p},\varphi) = -i \int^T_0\langle \Psi^{\text{GV}}_{{\bf p}}(t)|{\bf r}\cdot {\bf E}(t)|\Psi_0(t)\rangle  dt
\end {equation}
where  $T$ is the time duration of the laser pulse, $|\Psi_0(t)\rangle =|\psi_0\rangle \exp(iI_p t)$ is the ground state of an atom with ionization potential $I_p$,
\begin {equation}
|\Psi^{\text{GV}}_{{\bf p}}(t)\rangle= \exp({-i S_{{\bf p}}(t)})|{\bf p}+ {\bf A}(t)\rangle 
\end {equation}
is a Gordon–Volkov state \cite{Wolkow_1935} in length gauge, i.e., a solution of the time-dependent Schr\"odinger equation $i\partial_t|\Psi_{\bf p}^{\text{GV}}(t)\rangle = [{\bf p}^2/2 + {\bf r}\cdot{\bf E}(t)]|\Psi_{\bf p}^{\text{GV}}(t)\rangle$
without binding potential but laser only, so that
\begin{equation}\label{s}
S_{{\bf p}}(t)= \frac{1}{2}\int^t[{\bf p}+{\bf A}(t')]^2 dt'
\end{equation}
is the Coulomb-free, classical action of an electron in a laser field. We chose a hydrogen $1s$ state with $I_p = 0.5$ for $|\Psi_0(t)\rangle$. 

Figure~\ref{fig:fig1} shows a photoelectron spectrum \eqref{PES}, calculated with the SFA for a $20$-cycle, $\sin^2$-envelope laser pulse (\ref{A}) for $A_0=1$ ($E_0 = \omega = 0.0534$, corresponding to an intensity $I = 2\cdot 10^{14}~ \text{W/cm}^2$ and wavelength $\lambda = 854 ~\text{nm}$ of the main component), $\xi = 0.05$ (i.e., intensity ratio $I_{2\omega}/I_{\omega}=0.01$), and relative phase $\varphi = 0$. The three-fold symmetry of the photoelectron distribution coincides with the corresponding symmetry of the vector potential \footnote{The yield reaches its maxima  at times $t_k$ when the electric field peaks. The negative vector potential $-{\bf A}(t)$ at those emission times determines the maxima of the yield in the momentum plane. In our case we have $\omega t_k = (2k+1)\pi/3$, and the peaks of the yield occur under angles $\alpha_k = \omega t_k+\pi$ where the polar plot of ${\bf A}(t)$ peaks in magnitude, not $-{\bf A}(t)$ (as one might naively expect). Another example of such kind can be found in \cite{Mancuso_PhysRevA_2015}.}.

\begin{center}
    \includegraphics[width=0.4\textwidth]{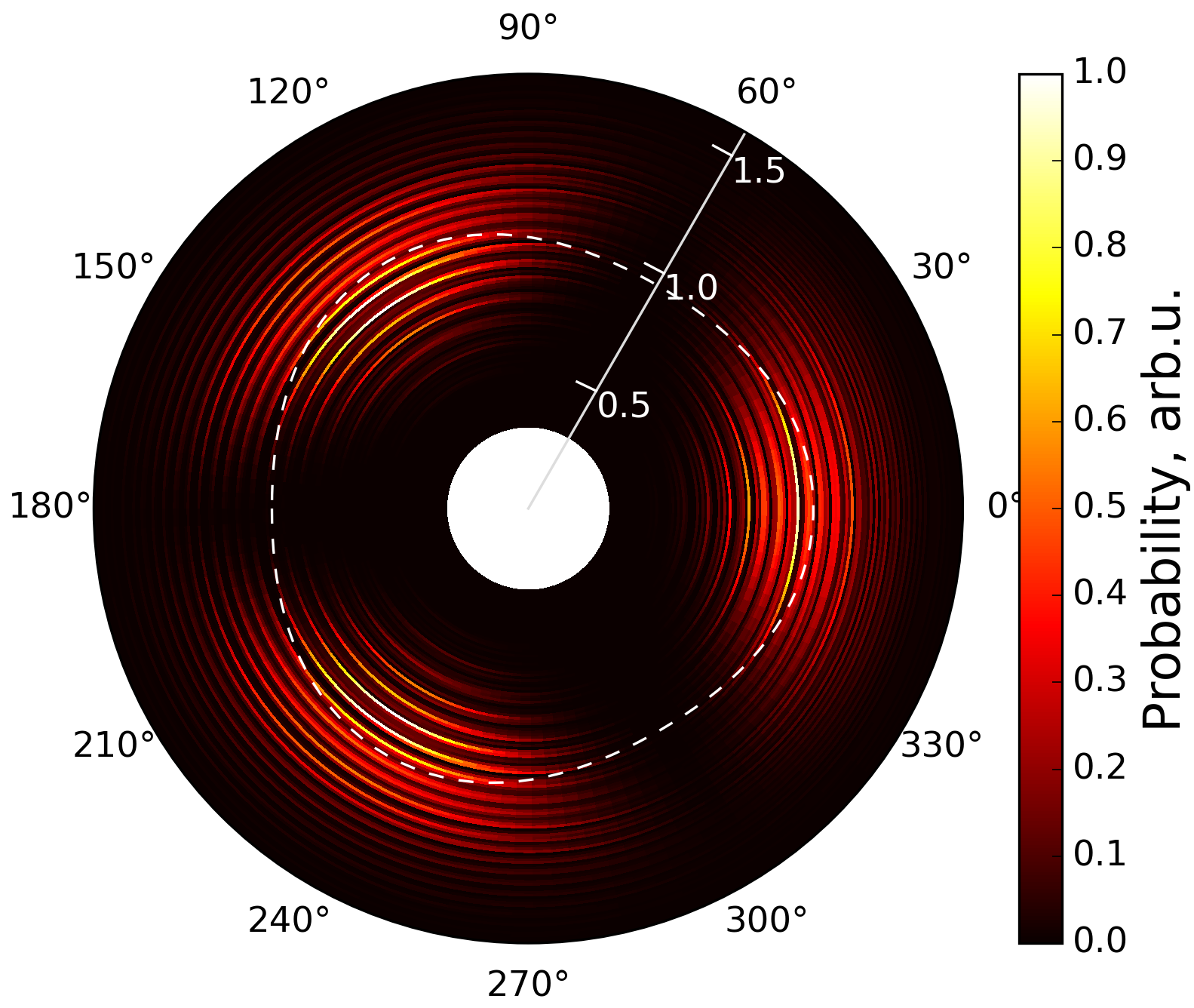} 
    \captionof{figure}{PES in the polarization plane, calculated with the SFA for the parameters stated in the text.  The time integral (\ref{Mpphi}) was calculated via the method of steepest descent \cite{Miloshevich_JPhysB_2006,Popruzhenko_JPhysB_2014}. The white, dashed line represents the vector potential ${\bf A}(t)$ during the central period of the laser pulse. The white axis indicates the radial momentum in atomic units.}
    \label{fig:fig1}
\end{center}

After the calculation of PES $Y({\bf p},\varphi)$ for a set of equally spaced phases $\varphi\in[0,2\pi[$ (typically ten) we Fourier-transform  them to extract the PP $\Phi_1({\bf p})$  \cite{Skruszewicz_PhysRevLett_2015, Almajid_JPhysB_2017}. The result is shown in Fig.~\ref{fig:fig2}. Besides the expected three-fold symmetry another pronounced feature is observed: the PP flips by a value of $\pi$ at a certain radial momentum $\simeq 1.7$. This radial momentum does, in general, not coincide with the maximum yield (see Fig.~\ref{fig:fig1}).   

\begin{center}
    \includegraphics[width=0.4\textwidth]{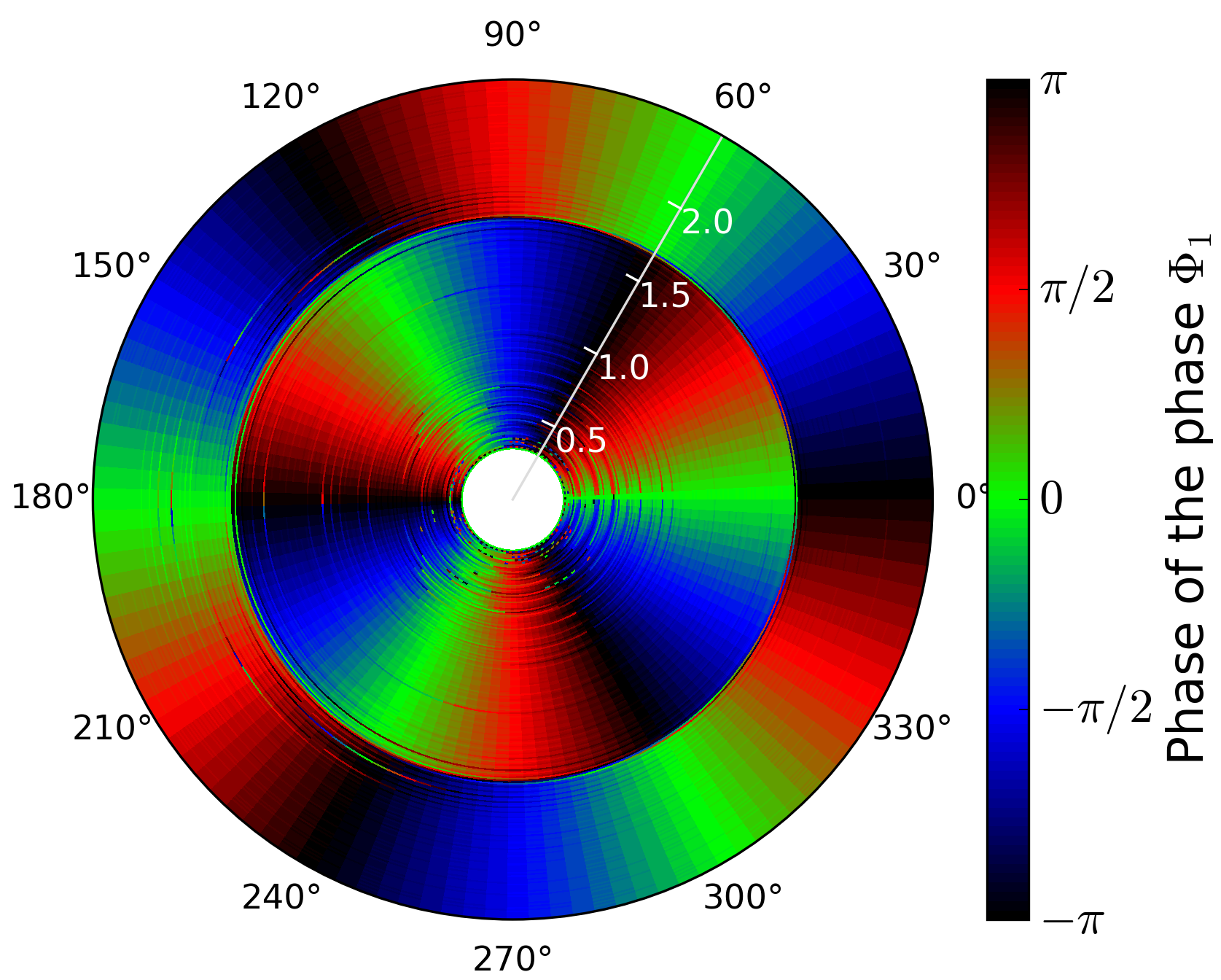}
    \captionof{figure}{The PP $\Phi_1({\bf p})$, calculated from the SFA-PES like the one shown in Fig.~\ref{fig:fig1} for ten values of $\varphi\in[0,2\pi[$.}
    \label{fig:fig2}
\end{center}

An analytical expression for the radial flipping momentum can be derived for the case of a long, flat-top pulse (i.e., $f(t)=1$, switched on and off at $\mp \infty$, respectively). In this case one may use the limit for the ionization rate
\begin{align}
\partial_t Y({\bf p},\varphi) & \simeq \lim_{T \rightarrow \infty} \frac{\vert M({\bf p},\varphi)\vert^2}{T}\nonumber \\ 
&\simeq \left\vert\int_0^{2\pi/ \omega} P({\bf p},t) \exp(-i S({\bf p},t))dt\right\vert^2\label{Y_inf}
\end{align}
where the exponent with the action
\beq
S({\bf p},t) = \int_t^{\infty} \left( \frac{({\bf p}+{\bf A}(t'))^2}{2}+I_p \right)dt'
\eeq
is a rapidly oscillating function, and $P({\bf p},t)$ is a slowly varying prefactor. The time integral in (\ref{Y_inf}) can be calculated with exponential accuracy by the method of steepest descent with complex-valued saddle points $t_s$ being solutions of the saddle point equation  \cite{Miloshevich_JPhysB_2006,Popruzhenko_JPhysB_2014}
\begin{equation}\label{sp0}
\left[{\bf p} + {\bf A}(t_s)\right]^{2} + 2 I_p=0.
\end{equation}
Writing  
\begin{equation}
 {\bf p} = p\cos\alpha ~{\bf e}_x + p\sin\alpha~ {\bf e}_y,
\end{equation}
where $\alpha$ is an angle of photoelectron emission in the polarization plane, one obtains from (\ref{sp0})
\begin{align}
p^2 +2 p A_0 \Big[\cos(\omega t_s-\alpha) +\xi \cos(2\omega t_s + \alpha+\varphi )\Big] & \nonumber \\
 +A_0^2 \Big[2\xi  \cos(3\omega t_s + \varphi ) + 1 +\xi^2\Big] + 2 I_p &= 0.
\end{align}
It is convenient to introduce the angle 
\begin{equation}\label{theta}
\theta = \varphi-3\alpha
\end{equation}
and dimensionless time, momentum and ionization energy (i.e., squared Keldysh parameter $\gamma$) as
\begin{equation}
\tau =\omega t_s-\alpha, \qquad q= \frac{p}{A_{0}}, \qquad \gamma^2 = \frac{2 I_p}{A_{0}^2}.
\end{equation}
The saddle point equation then boils down to
\begin{align}
q^2 + 2q \Big[\cos\tau +\xi \cos(2\tau+\theta)\Big] & \nonumber \\
 + 2 \xi \cos(3\tau+\theta) +1+\xi^2+\gamma^2 &= 0. \label{sp1}
\end{align}
The imaginary part of the action
\begin{align}
\text{Im}S(q, \tau) &= -\frac{A_{0}^2}{2\omega}  \text{Im}\bigg\{\Big(q^2 +1 +\gamma^2\Big) \tau + 2q \sin \tau \nonumber \\
& \quad +  \xi \Big[ q\sin (2 \tau +\theta )  + \frac{2}{3}  \sin (3\tau + \theta )\Big]\bigg\}
\end{align}
governs the ionization probability. In the monochromatic case ($\xi=0$) the solution to (\ref{sp1}) is \cite{PPT_1966,Smirnova_PhysRevA_2011}
\beq
 \tau^{0}  =  \pi+i\cosh^{-1}\bigg(\frac{1 + \gamma^2 + q^2}{2q}\bigg)
\eeq
so that, up to first order in $\xi$,
\begin{align}
\text{Im} S(q,\tau) & \simeq - \frac{A_{0}^2}{2\omega}  \bigg\{\Big(q^2 +1 +\gamma^2\Big)\tau^{0}_{i} - 2q\sinh\tau^{0}_{i} \nonumber \\
& + \xi\cos\theta\Big[q\sinh (2\tau^{0}_{i} )  - \frac{2}{3}\sinh (3\tau^{0}_{i})\Big]\bigg\}  \label{action1}
\end{align}
(the correction of order $\xi$ to the saddle point solution does not contribute in first order to the action). The term proportional to $\cos\theta$ in (\ref{action1}) changes sign (i.e., the phase jumps by $\pi$) when the square bracket vanishes,
\begin{equation}\label{flip}
q\sinh (2\tau^{0}_{i} )  - \frac{2}{3}\sinh (3\tau^{0}_{i}) = 0.
\end{equation}
For the parameters considered above this equation is fulfilled for the (dimensionless) radial momentum $q=1.68$. Because we chose $A_0=1$ in the example above, the value of the dimensionless momentum equals the momentum in atomic units. Hence we find excellent agreement with the position of the phase flip observed in Fig.~\ref{fig:fig2}.

\begin{center}
   \includegraphics[width=0.4\textwidth]{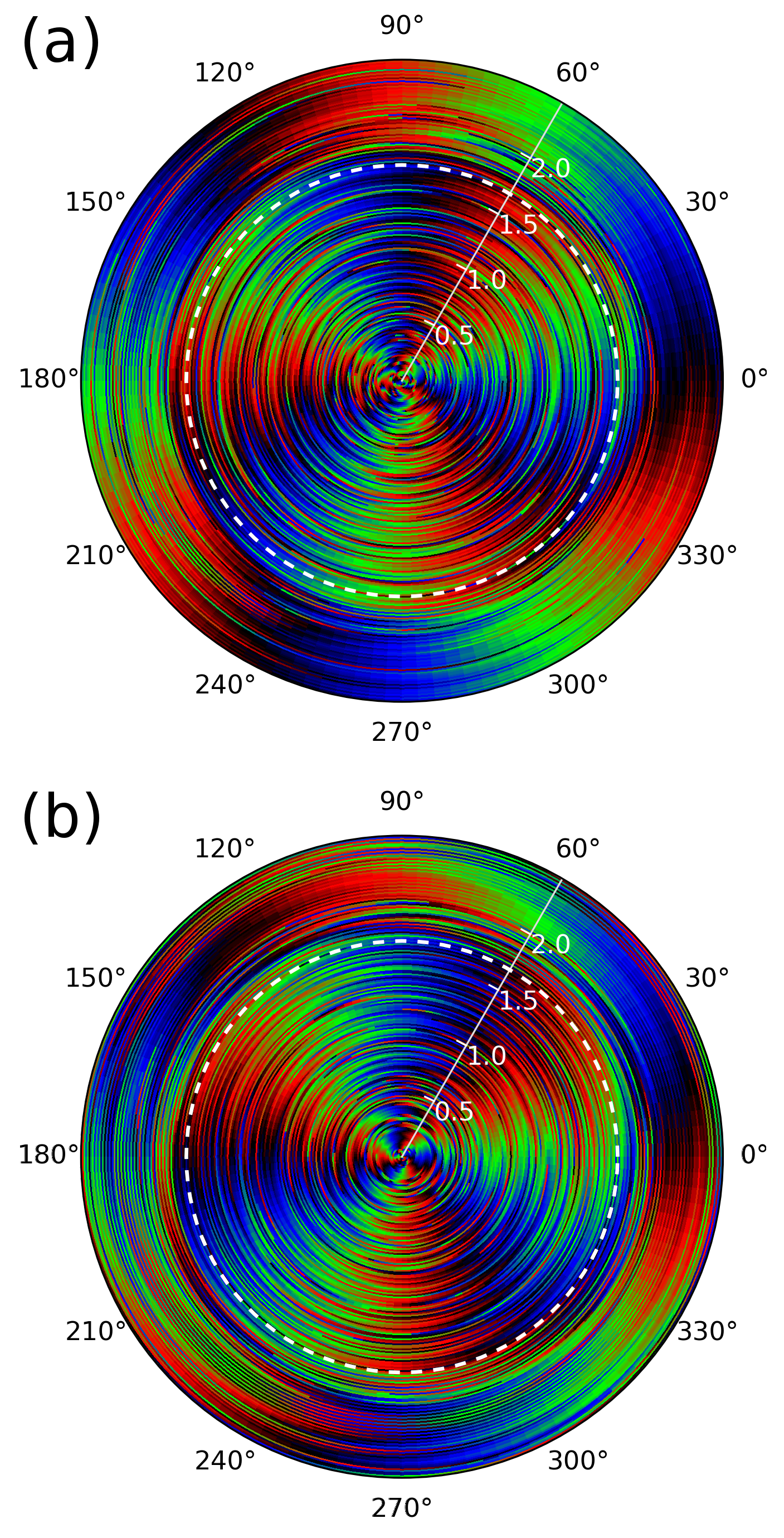}
   \captionof{figure}{(a) PP spectra obtained from the numerical solution of the TDSE for hydrogen initially in the $1s$ state for the same laser parameters as in the previous figures. (b) Same but for Coulomb potential cut at $r=1$ and $Z$ adjusted to reproduce the hydrogen $1s$ ionization potential. Dashed white lines indicate  the SFA-predicted position of the PP flip at $p=1.68$. The colormap is the same as in Fig.~\ref{fig:fig2}.}
    \label{fig:fig3}
\end{center}

In order to confirm our SFA results we compare them with results from the numerical \textit{ab initio} solution of the time-dependent Schr\"odinger equation (TDSE). The TDSE was solved for the hydrogen atom in the above specified laser field. We used Qprop \cite{qprop, qprop_tsurff}, which allows---within the dipole approximation---for arbitrary vector potentials in the $xy$-plane. The result for the PP spectrum is shown in Fig~\ref{fig:fig3}(a). The position of the PP flip appears to be in good agreement with the SFA prediction. An overall rotation in the low-energy region of the PP spectrum as compared to the SFA result is a well-known Coulomb effect \cite{Goreslavski_PhysRevLett_2004, Popruzhenko_PhysRevA_2008, Mancuso_PhysRevA_2015, Torlina_NaturePhys_2015}: apart from the spiraling in the laser field, the outgoing electron does not fly straight to the detector but on a curved, parabolic orbit due to the long-range Coulomb interaction with its parent ion. We additionally solved the TDSE with the Coulomb potential $-Z/r$ cut at $r=1$ and $Z$ adjusted to have the same ionization potential $0.5$ as the $1s$ state of hydrogen. In this case, no rotation of the PP spectrum is expected and, indeed, confirmed in Fig.~\ref{fig:fig3}(b). The PP flip is not much affected by the long-range Coulomb potential and close to the radial momentum predicted by the SFA. 

\section{CONCLUSION AND OUTLOOK} \label{concl}

An analytical expression for the phase-of-the-phase spectrum in a two-color ($\omega$-$2\omega$), circularly polarized, counter-rotating laser field was derived within the strong-field approximation. The phase of the phase is shown to have a well-distinguishable flip by $\pi$ at a certain radial photoelectron momentum. The prediction by the strong-field approximation is compared with the corresponding result from the numerical \textit{ab initio} solution of the time-dependent Schrödinger equation. Apart from an overall rotation of the spectrum due to the long-range Coulomb interaction between the outgoing electron and its parent ion (taken into account in the numerical solution but not in the strong-field approximation), the radial momenta at which the phase flip occurs are in good agreement. Because of its sharpness, the phase flip is a robust signature that should be easily detectable experimentally. On the other hand, its position is very sensitive to laser and target parameters such that it might be used for imaging the target (if the laser field is known) or the laser field (if the target is known). Another possible application is the extraction of useful information about the coherent ionization dynamics in complex many-body systems where photoelectron spectra might be cluttered by thermal, delayed or multiply scattered electrons.

\section*{Acknowledgment}
This work was supported by the project BA 2190/10 of the German Science Foundation (DFG).

\bibliography{biblio}
\end{document}